\documentclass[letter,twocolumn]{jpsj2} % two-column layout

\title{Indirect and direct energy gaps in the 
Kondo semiconductor YbB$_{12}$. } 

\author{%
Hidekazu \textsc{Okamura}\thanks{E-mail: okamura@kobe-u.ac.jp}, 
Takahiro \textsc{Michizawa}, 
Takao \textsc{Nanba}, 
Shin-ichi \textsc{Kimura}$^1$, 
Fumitoshi \textsc{Iga}$^2$, and 
Toshiro \textsc{Takabatake}$^2$
}

\inst{%
Graduate School of Science and Technology, Kobe University, 
Kobe 657-8501 
\\
$^1$UVSOR Facility, Institute for Molecular Science, 
Okazaki 444-8585. 
\\
$^2$ADSM, Hiroshima University, Higashi-Hiroshima 739-8530. 
}

\recdate{\today}

\abst{
Optical conductivity [$\sigma(\omega)$] of the Kondo 
semiconductor YbB$_{12}$ has been measured over wide 
ranges of temperature ($T$=8$-$690~K) and 
photon energy ($\hbar \omega \geq$ 1.3~meV).       
The $\sigma(\omega)$ data reveal the entire crossover 
of YbB$_{12}$ from a metallic electronic structure 
at high $T$ into a semiconducting one at low $T$.        
Associated with the gap development in $\sigma(\omega)$, 
a clear onset is newly found at $\hbar\omega$=15~meV 
for $T \leq$ 20~K.    
The onset energy is identified as the gap width 
of YbB$_{12}$ appearing in $\sigma(\omega)$.     
This gap in $\sigma(\omega)$ is interpreted 
as the indirect gap, which has been predicted in 
the band model of Kondo semiconductor.     
On the other hand, the strong mid-infrared (mIR) peak 
observed in $\sigma(\omega)$ is interpreted as arising 
from the direct gap.    
The absorption coefficient around the onset and the mIR peak 
indeed show characteristic energy dependences expected for 
indirect and direct optical transitions in conventional 
semiconductors.   

}

\kword{
YbB$_{12}$, Kondo semiconductor, optical conductivity, 
indirect gap}

\begin{document}
\maketitle

YbB$_{12}$ has been well known\cite{iga1,iga2,takaba} as an 
example of the Kondo semiconductors.\cite{takaba,fisk,riseborough}    
It develops a small energy gap at the Fermi level 
below 80~K.\cite{iga1,iga2}    
The gap width in YbB$_{12}$ has been estimated to be 
12.7~meV (136~K) from electrical 
resistivity,\cite{iga2} 15~meV (180~K) from Hall 
effect,\cite{iga2} and approximately 15~meV from 
electronic specific heat\cite{hiura} and 
photoemission\cite{takeda} experiments.     The gap 
formation mechanism in the Kondo semiconductor 
has been discussed extensively.\cite{takaba,fisk,riseborough}      
In the so-called band model, the gap is regarded as 
a band gap resulting from the hybridization between a wide 
conduction ($c$) band and a flat $f$-electron 
band.\cite{fisk,riseborough}     
Here the gap width is renormalized to a much smaller value 
than those in conventional semiconductors, due to strong 
correlation of the $f$ electrons.      To calculate the 
temperature ($T$) dependence of the energy gap 
based on the band model, the periodic Anderson 
Hamiltonian\cite{hewson} has been often used in 
conjunction with the dynamical mean field 
approximation.\cite{mutou,jarrel,rozenberg,logan}      
Band calculations of YbB$_{12}$ including both the 
orbital degeneracy and the spin-orbit coupling have 
been also reported.\cite{saso-harima,antonov}     
Recently, a theory of $T$-dependent energy gap in 
YbB$_{12}$ using realistic band structures has been 
reported.\cite{saso}

Previously, we reported an optical study of YbB$_{12}$ single 
crystals.\cite{okamura1}    The optical 
conductivity spectrum, $\sigma(\omega)$, of YbB$_{12}$ clearly 
showed an energy gap formation below 80~K.   The gap 
development involved a progressive depletion of 
$\sigma(\omega)$ below a shoulder at $\sim$ 40~meV.    
In addition, we observed a strong mid-infrared (mIR) 
absorption in $\sigma(\omega)$ peaked at $\sim$ 0.25~eV, 
which was also strongly $T$-dependent.      
The evolution of $\sigma(\omega)$ in the diluted system 
Yb$_{1-x}$Lu$_x$B$_{12}$ was also studied in 
detail.\cite{okamura2}

This Letter reports $\sigma(\omega)$ data measured 
on YbB$_{12}$ single crystals having a higher quality 
than the previous ones.\cite{okamura1}       
They show a residual resistivity ratio exceeding $10^5$ 
and no impurity-related Curie tail in the low-$T$ magnetic 
susceptibility.\cite{hiura}      
In addition, the temperature and photon energy ranges 
of the experiment have been extended, from $T$=20$-$290~K and 
$\hbar \omega$ $\geq$ 7~meV in the previous work\cite{okamura1} 
to $T$=8$-$690~K and $\hbar \omega$ $\geq$ 1.3~meV 
in this work.    
The obtained $\sigma(\omega)$ reveals the entire evolution 
of electronic structures from metallic to semiconducting 
ones with decreasing $T$.     Below 20~K, 
$\sigma(\omega)$ has revealed a clear onset at 15~meV, 
which we identify as the energy gap width.   The energy of 
15~meV agrees well with the gap widths obtained by other 
experimental techniques.\cite{iga2,takaba,hiura,takeda}    
We conclude that the observed energy gap of 15~meV in 
$\sigma(\omega)$ arises from an indirect gap, predicted 
by the band model of the Kondo semiconductor.   
The mIR peak, on the other hand, is interpreted as arising 
from the direct gap.    These assignments are supported 
by the analysis of absorption coefficient.

The single crystal of YbB$_{12}$ (batch number \#71) was grown 
as previously described.\cite{iga2}    A disk-shaped sample of 
4.5~mm diameter was cut from the crystal, and mechanically 
polished for optical measurements.     
The reflectivity spectrum, $R(\omega)$, of the sample was 
measured at the above-mentioned ranges of $T$ and 
$\omega$.\cite{footnote1}    
$\sigma(\omega)$ spectra were obtained from the 
Kramers-Kronig (K-K) analysis\cite{dressel} of the measured 
$R(\omega)$ spectra, which were extrapolated using the 
Hagen-Rubens function\cite{dressel} below the measured 
energy range.       
Other details of the optical experiments for 
$\hbar \omega$ $\geq$ 
4~meV were similar to those previously 
described.\cite{okamura1,okamura2}     
$R(\omega)$ between 1.3 and 4~meV (10-30~cm$^{-1}$) were 
measured using a THz synchrotron radiation source at the 
beam line BL6B of UVSOR,\cite{bl6b} Institute for 
Molecular Science.     
The sample was cooled by a continuous-flow liquid He 
cryostat.      A rapid-scan, Martin-Pupplet-type 
interferometer was used with an InSb bolometer to 
record $R(\omega)$.

Figures~1(a) and 1(b) show the measured $T$ dependences of 
$R(\omega)$ and $\sigma(\omega)$ in the infrared region.    
(Higher-energy spectra were reported previously.\cite{okamura1})  
With decreasing $T$ from 295~K, the broad dip centered 
near 0.15~eV in $R(\omega)$ becomes deeper, and $R(\omega)$ 
below 40~meV is strongly reduced at $T \leq$ 80~K.     
These two features in $R(\omega)$ give rise to the mIR peak 
in $\sigma(\omega)$ centered at 0.2-0.25~eV and the 
energy gap formation in $\sigma(\omega)$ below 40~meV.        
These overall features are very similar to those 
previously reported.\cite{okamura1}      
With increasing $T$ above 295~K, the mIR peak in 
$\sigma(\omega)$ becomes progressively weaker.      
At 690~K, the overall $\sigma(\omega)$ is similar 
to that of a metal, i.e., $\sigma(\omega)$ is basically 
a decreasing function of energy.    
Namely, the entire crossover from metallic to semiconducting 
(insulating) electronic structures in YbB$_{12}$ has 
been revealed in the present $\sigma(\omega)$ data.

The detailed $T$-evolutions of $R(\omega)$ and 
$\sigma(\omega)$ due to the gap formation are shown 
in Figs.~1(c) and 1(d), respectively.     
As in the previous result,\cite{okamura1} the gap formation 
in $\sigma(\omega)$ starts around 80~K, with a shoulder 
appearing at $\sim$ 40~meV (indicated by the black arrow).   
As the gap develops with decreasing $T$, the density 
of free carriers decreases progressively.\cite{iga2}     
This results in, as seen in Figs.~1(c) and (d), 
the shifts of the plasma edge (sharp minimum) 
in $R(\omega)$ and the decrease of the Drude-like 
component (the rise toward $\omega$=0) in 
$\sigma(\omega)$.      
At 20 and 8~K, $\sigma(\omega)$ has a clear onset at 
15~meV (indicated by the red arrow), below which the 
spectral weight is very small.     
The onset in $\sigma(\omega)$ results from the hump in 
$R(\omega)$ located around 15~meV (indicated by the 
blue arrow).   
This hump in $R(\omega)$, which had not been observed 
previously,\cite{okamura1} has been reproduced in 
repeated experiments.     
Although $R(\omega)$ at 8~K does not approach 1 even near 
$\omega$=0, a Hagen-Rubens extrapolation has been used 
similarly to other data at higer $T$'s, since a small 
density of free carriers are still present even at 
8~K.\cite{iga2,hiura,plasma}    
We have found, however, that the onset of $\sigma(\omega)$ 
at 15~meV is unaffected by details of the extrapolation used.   
Hence, the onset is neither due to an experimental error 
nor due to an artifact of the K-K analysis, but it should 
be an intrinsic feature.    We conclude that the energy 
gap width of YbB$_{12}$ observed in $\sigma(\omega)$ is 
15~meV, as given by the onset energy.      
We had previously overestimated the gap width to be 25~meV, 
due to the lack of a clear onset in the previous 
$\sigma(\omega)$ data.\cite{okamura1}     
Compared with the above-mentioned gap widths ($E_g$) 
in the total density of states (DOS) obtained by other 
experimental techniques,\cite{iga2,iga3,takeda} it is 
clear that the gap width in $\sigma(\omega)$ is just 
equal to $E_g$.      
It is interesting that a magnetic excitation peak has 
also been observed at $\sim$ 15~meV in the inelastic 
neutron scattering of YbB$_{12}$.\cite{neutron1,neutron2}

According to the band model of Kondo semiconductor, 
as sketched in Fig.~2, the $c$-$f$ hybridization state 
involves both an indirect gap and a direct 
gap.\cite{fisk,riseborough}     
The indirect gap corresponds to $E_g$, which can be 
measured by various experiments as already mentioned.      
Hence, in the band model of YbB$_{12}$, the present 
result indicates that {\it the energy 
gap appearing in $\sigma(\omega)$ is the indirect gap}.     
In fact, the band calculation has shown that the minimum 
energy gap in YbB$_{12}$ should be an indirect 
gap.\cite{saso-harima}       
Note that, however, such indirect optical transitions 
are forbidden by the momentum conservation rule within 
the first-order optical processes, which involve the 
absorption of a photon only.\cite{dressel,cardona}     
This point will be discussed later.

We had previously conjectured\cite{okamura1} that the 
gap in $\sigma(\omega)$ corresponded to the direct gap and the 
mIR peak to some interband transition due to the 
band structure of YbB$_{12}$.     
Since then, however, YbAl$_3$, a typical mixed-valent metal, 
has also shown a mIR peak\cite{okamura3} very similar to that 
for YbB$_{12}$.    
In addition, many other Yb-based metals, including 
YbInCu$_4$,\cite{schle} YbRh$_2$Si$_2$,\cite{kimura} 
YbAl$_2$, YbCu$_2$Si$_2$, YbNi$_2$Ge$_2$ and 
YbCuAl,\cite{okamura4} have also shown a 
similar mIR peak.      
Namely, {\it the mIR peak is a universal feature for 
these Yb-based compounds}, and it is likely to result 
from a common electronic structure shared by these 
compounds.       
Within the band model of a mixed-valent or a 
heavy-fermion metal,\cite{hewson} the underlying 
electronic structure is exactly analogous to that for 
the Kondo semiconductor sketched in Fig.~2.    
Hence, these results strongly suggest that the 
{\it mIR peak is due to the optical transitions 
across the direct gap in the band model}, 
as sketched in Fig.~2.

We have interpreted the onset of $\sigma(\omega)$ at 15~meV 
as the absorption edge at the indirect gap and the mIR peak 
as arising from the direct gap within the band model.    
As already mentioned, indirect optical 
transitions are forbidden within the first-order optical 
processes.      However, they are allowed within the second-order 
processes involving the absorption/emission of 
a photon {\it and a phonon}, the latter of which provides 
the required momentum transfer.\cite{cardona}    
In such an indirect transition, a phonon is either emitted 
or absorbed by the electron, followed by or preceded by a 
virtual electronic transition to or from an intermediate 
state.     
The process with a phonon absorption can occur only at 
sufficiently high $T$, but that with a phonon emission 
may occur even at low $T$.       
It is well known that indirect-gap semiconductors actually 
show phonon-assisted indirect absorption even at 
low $T$.\cite{cardona}      
A good example is Ge,\cite{Ge} which is a band semiconductor 
with the indirect gap ($E_{ind}$) of $\sim$ 0.74~eV and the 
minimum direct gap ($E_{dir}$) of $\sim$ 0.89~eV at 4.2~K.    
At 4.2~K, Ge shows quite sizable absorption at photon 
energies well below $E_{dir}$.\cite{Ge}    
The photon-energy dependence of the optical absorption 
coefficient\cite{absorption} ($\alpha$) near 
the direct and indirect absorption edges 
can be calculated assuming parabolic band edges: 
$\alpha^2$ is linear in energy near the direct one 
and $\alpha^\frac{1}{2}$ is linear in energy near 
the indirect one, where the energy is measured from 
$E_{dir}$ and $E_{ind}$, respectively.\cite{cardona}     
Such dependences have been actually observed 
for conventional semiconductors.\cite{cardona,Ge}    
For comparison, $\alpha^2$ and $\alpha^\frac{1}{2}$ 
of YbB$_{12}$, obtained from the K-K analysis of $R(\omega)$, 
are plotted in Fig.~3 together with $\sigma(\omega)$.        
$\alpha^2$ is apparently linear in energy over a width 
of $\sim$ 0.06~eV, which coincides with the range where 
the mIR peak is centered in $\sigma(\omega)$.     
$\alpha^\frac{1}{2}$ is also linear in energy above the 
onset at 15~meV.      
Hence, the present optical data possess the well-known 
characteristics of indirect and direct 
absorptions for conventional semiconductors.     
The intercept of the linear portion (the broken line 
in Fig.~3) 
with the energy axis gives the gap width.\cite{cardona}      
Hence the direct gap in YbB$_{12}$ is estimated to be 
$\sim$ 0.18~eV in this analysis.

As mentioned above, the characteristic energy dependence 
of $\alpha$ for indirect and direct transitions have been 
derived for simple parabolic bands.\cite{cardona}     
The actual band 
dispersions in YbB$_{12}$ should be far more complicated, 
as actually shown by the band 
calculations.\cite{antonov,saso-harima}      
In addition, it is even unclear whether phonon-assisted 
indirect transitions are indeed possible in YbB$_{12}$, 
since such a process has not been examined with the 
phonon dispersion of YbB$_{12}$ taken into account.     
Therefore, it is rather surprising that the measured 
$\alpha$ of YbB$_{12}$ closely follows the characteristic 
energy dependence predicted by the simplified model.

Theoretical calculations of $T$-dependent $\sigma(\omega)$ 
for the Kondo semiconductor have been made using the periodic 
Anderson Hamiltonian with the dynamical mean field 
approximation.\cite{mutou,jarrel,rozenberg,logan}     
Earlier works\cite{mutou,jarrel,rozenberg} assumed first 
order optical processes, hence the calculated $\sigma(\omega)$ 
contained the direct gap only.   Accordingly, the calculated 
$\sigma(\omega)$ at low $T$ showed an extremely sharp onset 
at $E_{dir}$.   Hence, to compare the calculated 
$\sigma(\omega)$ with the experimental $\sigma(\omega)$, 
some broadening would have to be introduced into the former.      
Recently, it has been suggested\cite{logan} that many-body 
scattering, inherent in strongly correlated $f$-electron system, 
may provide the momentum transfer needed for indirect 
transitions.   
The calculated $\sigma(\omega)$ in this model\cite{logan} 
showed a long tail below $E_{dir}$, showing better 
agreement with the experimental $\sigma(\omega)$.     
Note that such a process is analogous to the 
phonon-assisted one, the difference being the 
source of the extra momentum.       
Since the appearance of indirect transitions in the measured 
$\sigma(\omega)$ is not likely to result from simple 
broadening effects, it is important to further examine 
these processes.    
In the theory by Saso,\cite{saso} both indirect and direct 
gaps in the band structure of YbB$_{12}$ were included to 
calculate the energy gap in $\sigma(\omega)$.     This approach was 
taken because retaining only the direct gaps in the 
realistic band structure of YbB$_{12}$ gave an energy gap 
in $\sigma(\omega)$ much larger than the observed one.\cite{saso}      
We propose that the combination of phonon-assisted 
indirect transitions with the direct gaps in the band 
structure may actually improve the agreement with 
the experiment.

In conclusion, the present $\sigma(\omega)$ data reveal 
the entire crossover of YbB$_{12}$ from metallic to 
semiconducting electronic structures.    
At low $T$, a clear onset of $\sigma(\omega)$ is 
observed at 15~meV, 
and this energy is identified as the gap width 
in $\sigma(\omega)$.    
This value agrees very well with $E_g$ in the total 
DOS obtained by other experiments.   
The observed gap in $\sigma(\omega)$ is interpreted 
as the indirect gap within the hybridization-band model 
of the Kondo semiconductor.     
The energy dependence of the absorption coefficient 
around the onset and the mIR peak are indeed found 
to follow the characteristic patterns of indirect 
and direct absorptions, respectively, known for 
conventional band semiconductors.     
It is suggested that phonon-assisted indirect transitions 
may be important for a better understanding 
of $\sigma(\omega)$ for YbB$_{12}$.

We thank Dr. M. Matsunami for technical assistance 
in the high temperature measurements.   We also thank 
Prof. T. Saso and Prof. T. Mutou for useful discussions.    
This work has been partly supported by Grants-In-Aid 
(15654041 and 13CE2002) from the Ministry of Culture, 
Education, Sports, Science and Technology.      
A part of this work was performed as a Joint Studies 
Program of the Institute for Molecular Science (2004).

\pagebreak

%%%%%%%%%%%%%%%%%%%%%%  FIG.1  %%%%%%%%%%%%%%%%%%%
\begin{figure}
\begin{center}
\caption{(Color) 
(a) Optical reflectivity $R(\omega)$ and (b) conductivity 
$\sigma(\omega)$ of YbB$_{12}$ between 8 and 690~K.     
The weak dip in $R(\omega)$ around 0.4~eV at 8~K is due 
to the absorption by ice formed on the sample, which has 
negligible effect on $\sigma(\omega)$ in (b).     
(c) and (d) show $R(\omega)$ and $\sigma(\omega)$, 
respectively, below 120~K in the low-energy region.    
The blue arrow in (c) indicates the hump in $R(\omega)$, 
and the black and red arrows in (d) indicate, 
respectively, the shoulder and the onset in $\sigma(\omega)$.   
The broken curves are the extrapolations (see text).   

}
\end{center}
\end{figure}
%%%%%%%%%%%%%%%%%%%%%%%%%%%%%%%%%%%%%%%%%%%%%%%%%%%

%%%%%%%%%%%%   FIG.2   %%%%%%%%%%%%%%
\begin{figure}
\begin{center}
\caption{
Sketch of the band model for the Kondo 
semiconductor.\cite{fisk,riseborough}  The hybridization 
between an otherwise flat $f$ band and a wide conduction 
band results in the direct gap and the indirect gap 
($E_g$).    The arrows indicate the relevant optical 
transitions.    
}
\end{center}
\end{figure}

%%%%%%%%%%%%   FIG.3   %%%%%%%%%%%%%%
\begin{figure}
\begin{center}
\caption{
(a) Optical conductivity $\sigma$, (b) absorption 
coefficient $\alpha$, (c) $\alpha^2$, and 
(d) $\alpha^\frac{1}{2}$ of YbB$_{12}$ at 8~K 
as a function of photon energy.   
The broken lines are guide to the eye, indicating 
the linear dependence on photon energy discussed 
in the text.    
}
\end{center}
\end{figure}


\begin{thebibliography}{99}

\bibitem{iga1} M. Kasaya, F. Iga, K. Negishi, S. Nakai 
and T. Kasuya: 
J. Magn. Magn. Mater. {\bf 31-34} (1983) 437. 

\bibitem{iga2} F. Iga, N. Shimizu and T. Takabatake: 
J. Magn. Magn. Mater. {\bf 177-181} (1998) 337.    

\bibitem{takaba} T. Takabatake, F. Iga, T. Yoshino, 
Y. Echizen, K. Katoh, K. Kobayashi, M. Higa, N. Shimizu, 
Y. Bando, G. Nakamoto, H. Fujii, K. Izawa, T. Suzuki, 
T. Fujita, M. Sera, M. Hiroi, K. Maezawa, S. Mock, 
H.v. Lohneysen, A. Bruckl, K. Neumaier and K. Andres: 
J. Magn. Magn. Mater. {\bf 177-181} (1998) 277.

\bibitem{fisk} G. Aeppli and Z. Fisk: 
Comments Condens. Matter Phys. {\bf 16} (1992) 155. 

\bibitem{riseborough} P.S. Riseborough: 
Adv. Phys. {\bf 49} (2000) 257.   

\bibitem{hiura} S. Hiura: Master Thesis, Hiroshima 
University (2000).  

\bibitem{takeda} Y. Takeda, M. Arita, M. Higashiguchi, 
K. Shimada, M. Sawada, H. Sato, M. Nakatake, H. Namatame, 
M. Taniguchi, F. Iga, T. Takabatake, K. Takata, E. Ikenaga, 
M. Yabashi, D. Miwa, Y. Nishino, K. Tamasaku, T. Ishikawa, 
S. Shin and K. Kobayashi: Physica B {\bf 351} (2004) 286.   

\bibitem{hewson} A.C. Hewson: {\it The Kondo Problem to Heavy 
Fermions} (Cambridge University Press, Cambridge, 1993). 

\bibitem{mutou} T. Mutou and D.S. Hirashima: 
J. Phys. Soc. Jpn. {\bf 63} (1994) 4475.  

\bibitem{jarrel} M. Jarrel: 
Phys. Rev. B {\bf 51} (1995) 7429.  

\bibitem{rozenberg} M.J. Rozenberg, G. Kotliar and H. Kajueter: 
Phys. Rev. B {\bf 54} (1996) 8452.   

\bibitem{logan} N.S. Vidhyadhiraja, V.E. Smith, D.E. Logan 
and H.R. Krishnamurthy: 
J. Phys. Condens. Matter {\textbf 15} (2003) 4045.   

\bibitem{saso-harima} T. Saso and H. Harima: 
J. Phys. Soc. Jpn. {\bf 72} (2003) 1131, 
and references therein.   

\bibitem{antonov} V.N. Antonov, B.N. Harmon and A.N. Yaresko: 
Phys. Rev. B {\bf 66} (2002) 165209.    

\bibitem{saso} T. Saso: 
J. Phys. Soc. Jpn. {\bf 73} (2004) 2894.  

\bibitem{okamura1} 
H. Okamura, S. Kimura, H. Shinozaki, T. Nanba, 
F. Iga, N. Shimizu and T. Takabatake: 
Phys. Rev. B {\bf 58} (1998) R7496.  

\bibitem{okamura2} 
H. Okamura, M. Matsunami, T. Inaoka, S. Kimura, 
T. Nanba, F. Iga, S. Hiura, J. Klijn and T. Takabatake: 
Phys. Rev. B {\bf 62} (2000) R13265.  

\bibitem{footnote1} 
$R(\omega)$ was measured at 0.0013$-$1~eV below 40~K, 
at 0.004$-$1~eV between 50 and 80~K, and at 0.01$-$1~eV 
between 80 and 295~K.     Obtained $R(\omega)$ spectra 
were smoothly connected to the previous ones above 
1~eV.\cite{okamura1}      Spectra above 295~K were 
measured at 0.055$-$1.8~eV, and smoothly connected 
to the room temperature spectrum above 1.8~eV.\cite{okamura1}   

\bibitem{dressel}
M. Dressel and G. Gruner: {\it Electrodynamics of Solids} 
(Cambridge University Press, Cambridge, 2002).  

\bibitem{bl6b} S. Kimura, E. Nakamura, J. Yamazaki, M. Katoh, 
T. Nishi, H. Okamura, M. Matsunami, L. Chen and T. Nanba: 
AIP Conf. Proc. {\bf 705} (2004) 416.    

\bibitem{plasma} The plasma edge in $R(\omega)$ has been 
observed down to $T$=12~K (not shown here) within our 
measurement range.    

\bibitem{neutron1} A. Bouvet, T. Kasuya, M. Bonnet, L.P. Regnault, 
J. Rossad-Mignot, F. Iga, B. Fak and A. Severing: 
J. Phys. Condens. Mat. {\bf 10} (1998) 5667.   

\bibitem{neutron2} E.V. Nefeodova, P.A. Alekseev, J.-M. Mignot, 
V.N. Lazukov, I.P. Sadikov, Yu. B. Paderno, N.Yu. Shitsevalova 
and R.S. Eccleston: Phys. Rev. B {\bf 60} (1999) 13507.  

\bibitem{cardona} P.Y. Yu and M. Cardona: {\it Fundamentals 
of Semiconductors} (Springer, Berlin, 2001) 3rd ed., 
Chapter 6.2.   

\bibitem{okamura3} H. Okamura, T. Michizawa, T. Nanba 
and T. Ebihara: 
J. Phys. Soc. Jpn. {\bf 73} (2004) 2045.  

\bibitem{schle} S.R. Garner, J.N. Hancock, Y.W. Rodriguez, 
Z. Schlesinger, B. Bucher, Z. Fisk and J.L. Sarrao: 
Phys. Rev. B {\bf 62} (2000) R4778.     

\bibitem{kimura} S. Kimura, T. Nishi, J. Sichelschmidt, 
V. Voevodin J. Ferstl, C. Geibel and F. Steglich: 
J. Magn. Magn. Mater. {\bf 272-276} (2004) 36.     

\bibitem{okamura4} H. Okamura, T. Watanabe, T. Nanba and 
N. Tsujii: Unpublished.     

\bibitem{Ge} G.G. MacFarlane, T.P. McLean, J.E. Quarrington 
and V. Roberts: Phys. Rev. {\bf 108} (1957) 1377. 

\bibitem{absorption} The optical absorption coefficient, $\alpha$, 
of a medium is defined as $I(z) = I(0) \exp(-\alpha z)$, 
where $I(z)$ is the light intensity within the medium at 
position $z$ along the direction of light propagation.   
$\alpha(\omega) = (2 \pi \omega / c)k(\omega)$, where $c$ is the 
speed of light and $k(\omega)$ is the imaginary part of the complex 
refractive index, hence $\alpha(\omega)$ can be calculated 
from the K-K analysis of $R(\omega)$.\cite{dressel,cardona}   



\end{thebibliography}
\end{document}